\documentclass[useAMS,usenatbib]{mn2e}

\usepackage{psfig,epsfig,supertabular,wrapfig,placeins}
\usepackage{graphicx,amsmath,amssymb,subfigure,multirow}
\usepackage{marvosym, times}
\usepackage{graphics}
\usepackage{natbib}
\usepackage{amssymb}
\usepackage{amsmath}
\usepackage{euscript}
\usepackage{setspace}
\usepackage{fancyhdr}
\usepackage{afterpage,rotating,url}

\newcommand{\etal}{{et al.}}                   

\newcommand{\Msolar}{\mbox{\,$\rm M_{\odot}$}}        
\newcommand{\Lsolar}{\mbox{\,$\rm L_{\odot}$}}        
 
\newcommand{\asec}{\ensuremath{^{\prime\prime}}}

\newcommand{\Oiii}{$[$O{\sc iii}$]$}
\newcommand{\Heii}{He{\sc ii}}
\newcommand{\Hbb}{H$\beta$b}
\newcommand{\fluxunits}{\ensuremath{{\rm~erg~cm}^{-2} {\rm ~s}^{-1}}}
\newcommand{\lumunits}{\ensuremath{{\rm ~erg~s}^{-1}}}
\newcommand{\gtsim}{\mbox{{\raisebox{-0.4ex}{$\stackrel{>}{{\scriptstyle\sim}}$}}}}
\newcommand{\ltsim}{\mbox{{\raisebox{-0.4ex}{$\stackrel{<}{{\scriptstyle\sim}}$}}}}

\title[An 80-kpc Lyman-$\alpha$ halo around a Type-2 QSO]{An 80-kpc
  Lyman-$\alpha$ halo around a high-redshift type-2 QSO}

\author[D.J.B.~Smith, M.J.~Jarvis, C.~Simpson \& A.~Mart\'inez-Sansigre]{D.J.B. Smith$^{1,2}$\thanks{E-mail: djs@astro.livjm.ac.uk(DS)}, M.J.~Jarvis$^{3}$, C.~Simpson$^{1}$ \& A.~Mart\'inez-Sansigre$^4$ \\
    $^{1}$Astrophysics Research Institute, Liverpool John Moores University, Twelve Quays House, Egerton Wharf, Birkenhead, CH41 1LD, UK\\ 
    $^{2}$Department of Astrophysics, University of Oxford, Denys Wilkinson Building, Keble Road, Oxford, OX1 3RH, UK \\
    $^{3}$Centre for Astrophysics, Science \& Technology Research Institute, University of Hertfordshire, Hatfield, Herts, AL10 9AB, UK \\
    $^4$Max-Planck-Institut f\"ur Astronomie, K\"onigstuhl 17, D-69117 Heidelberg, Germany}
    
\begin{document}

\date{\today}

\pagerange{\pageref{firstpage}--\pageref{lastpage}} \pubyear{2008}

\maketitle

\label{firstpage}

\begin{abstract}
We announce the discovery of an extended emission line region
associated with a high redshift type-2 QSO. The halo, which was
discovered in our new wide-field narrow-band survey, resides at $z =
2.85$ in the Spitzer First Look Survey region and is extended over
$\sim$80 kpc. Deep VLBI observations imply that approximately 50 per
cent of the radio emission is extended on scales $>$ 200pc. The
inferred AGN luminosity is sufficient to ionize the extended halo, and
the optical emission is consistent with being triggered coevally with
the radio source. The Lyman-$\alpha$ halo is as luminous as those
found around high redshift radio galaxies, however the active nucleus
is several orders of magnitude less luminous at radio wavelengths than
those FRIIs more commonly associated with extended emission line
regions. AMS05 appears to be a high-redshift analogue to the
radio-quiet quasar E1821+643 which is core dominated but which also
exhibits extended FRI-like structure and contains an optically
powerful AGN. We also find evidence for more quiescent kinematics in
the Lyman-$\alpha$ emission line in the outer regions of the halo,
reminiscent of the haloes around the more powerful FRIIs.  The optical
to mid-infrared SED is well described by a combination of an obscured
QSO ($L_{\rm bol} \sim 3.4 \pm 0.2 \times 10^{13}$ \Lsolar) and a
1.4~Gyr old simple stellar population with mass $\sim 3.9 \pm 0.3
\times 10^{11}$~M$_{\odot}$.
\end{abstract}


\section{Introduction}
\label{intro}

\subsection{Type-2 QSOs}

Unified schemes rely on different lines of sight toward the active
galactic nucleus (AGN) to explain a variety of different species of
apparently distinct astrophysical phenomena (e.g. Antonucci \etal,
1993). These unified schemes invoke the putative dusty torus around
the central accreting supermassive black hole which obscures the
central nuclear source and the high-velocity gas which lies close to
the nucleus, i.e. the broad-line region. Although slight modifications
to this simple scheme may be needed (e.g. Lawrence 1991; Simpson 2005)
the general picture remains the same. The evidence for this
orientation-based unified scheme is now compelling, including the
properties of the unresolved hard X-ray background for example which
suggests the existence of a large number of obscured QSOs (e.g. Gilli
et al., 2007). The dusty torus is thought to obscure the central
engine along certain lines of sight, resulting in the nuclear source
appearing faint at optical wavelengths with the emission dominated by
the host galaxy's stellar population. Galaxies whose properties and
orientation causes us to observe them in this manner are known as
obscured (or type-2) QSOs.

Further evidence of these obscured sources comes from radio surveys,
where large-scale radio jets provide a method of finding both obscured
and unobscured AGN without the problems associated with dust
obscuration. These studies have shown that the fraction of obscured
(type-2) radio-loud AGN may be as high as 60 per cent, with the
unobscured (type-1) QSOs traditionally found in optical surveys only
comprising 40 per cent of the powerful AGN population (e.g. Willott et
al. 2000). Therefore, it would be reasonable to assume that a similar
fraction of radio-quiet obscured sources should also exist.  However,
such obscured sources have proven very difficult to find with
traditional techniques at optical wavelengths as they just appear as
normal galaxies in imaging data.

Although X-ray (Norman et al. 2002; Stern et al. 2002) and pure
spectroscopic (Jarvis et al. 2005) observations can find such sources
at high redshift, it is with the {\em Spitzer Space Telescope} that
this field has been revolutionized, with many groups using the fact
that the dusty torus, heated by the central AGN, reradiates at
mid-infrared wavelengths. Many methods based on mid-infrared
observations have been used to find such AGN; Lacy et al. (2004) and
Alonso-Herrero et al. (2006) for example used mid-infrared selection
criteria to identify them. The combination of radio and mid-infrared
data can also be used to reduce the fraction of star-burst dominated
sources from colour selected samples (e.g.  Donley et al. 2005,
Mart\'inez-Sansigre et al. 2005, 2006;~\S\ref{intro:ams05}), and deep
radio surveys have been shown to harbour a considerable number of
these sources (e.g. Simpson \etal, 2006).  All of these now seem to
imply that a large population of obscured AGN exist in the
high-redshift Universe and that many may also be missed in the current
deep X-ray surveys (e.g. Lacy et al. 2007; Mart\'inez-Sansigre et
al. 2007).

\subsection{Lyman-$\alpha$ Haloes}

In recent years much effort has been devoted to the study of high
redshift nebul\ae\ characterised by large luminosity and great extent
in the Lyman-$\alpha$ emission line -- typically L~$\sim~
10^{44}$\lumunits, extending over $\gtsim$50kpc. The existence of such
extended emission line regions (EELRs) around HzRGs had been known for
more than 10 years (e.g. McCarthy \etal, 1990) before the discovery of
two similar haloes at $z = 3.01$ by Steidel \etal\ (2000), which were
not associated with detections at radio wavelengths.

Many EELRs are now known to exist around HzRGs following the results
of e.g. Eales \etal\ (1993), Maxfield \etal\ (2002), Reuland
\etal\ (2003), Villar-Mart\'in \etal\ (2003), to name but a
handful. Lyman-$\alpha$ haloes have also been shown to reside around
QSOs by e.g. Bunker \etal\ (2003), Weidinger \etal\ (2005), and Barrio
\etal\ (2008). Furthermore, it seems that these extensive
Lyman-$\alpha$ haloes also exist around several galaxies similar to
those discovered by Steidel \etal\ (2000), which are not associated
with powerful AGN. Currently, tens of these galaxies are known
following the results of Francis \etal\ (2001), Matsuda \etal\ (2005),
Nilsson \etal\ (2006), and Smith \& Jarvis (2007). Results show that
the ionizing properties of optically-selected extended Lyman-$\alpha$
haloes are diverse, broadly falling into three categories (ionization
by AGN, galaxy wide super-wind, or a cold accretion process -- see
e.g. Smith \etal, 2008, for a more in depth discussion). Searches to
locate these huge line-emitting structures have generally been
serendipitous (e.g. Francis \etal, 2001, Dey \etal, 2005 -- shown to
harbour an obscured AGN within the halo -- or Nilsson \etal, 2006) or
targetted on known overdensities (e.g. Steidel \etal, 2000, Reuland
\etal, 2003, Matsuda \etal, 2005, Villar-Mart\'in \etal, 2003,2005).

Throughout this paper, the AB magnitude system is used (Oke \& Gunn,
1983), and a standard cosmology is assumed in which $H_{0}$ = 71 km
s$^{-1}$, $\Omega_{M}$ = 0.27 and $\Omega_{\Lambda}$ = 0.73 (Dunkley
\etal, 2008).

\section{Observations}\label{Observations}

\subsection{A new narrow-band survey}

In order to find the most luminous and extended Lyman-$\alpha$ haloes
and quantify their diverse ionization mechanisms, we conducted a
blank-field narrow-band survey with sufficient area to detect a large
sample of them at $z \sim 3$. We observed a total of 15~deg$^2$ using
the Wide Field Camera (WFC) on the Isaac Newton Telescope, and spread
our observations over three extragalactic fields (the XMM-LSS, Lockman
Hole, and Spitzer First Look Survey), enabling us to take advantage of
the wide range of high-quality public survey data available in these
regions of sky. We also split our observations between three different
narrow-band filters; \Heii$_{4686\slash 100}$, \Hbb$_{4861\slash 170}$, and
\Oiii$_{5008\slash 100}$. This improves the efficiency of our survey
strategy by reducing our requirements for Sloan-g' band observations
(i.e. each pointing requires observations in three separate
narrow-band filters, plus Sloan-g' band). Our typical 50\%
completeness limits in the narrow- and broad bands were nb$_{\rm AB}
\sim 24.5$ and g'$_{\rm AB} \sim 25.5$ respectively (measured in
4\asec\ apertures). The survey strategy, sensitivity and results will
be discussed in more detail in Smith \& Jarvis (\textit{in prep}).

Candidate Lyman-$\alpha$ emitters are separated from other objects
whose emission lines are redshifted into the narrow-band filter's
transmission function by virtue of their extremely high equivalent
widths (EWs). EWs are estimated based on our broad- and narrow-band
photometry (see e.g. Venemans \etal, 2005; Smith \& Jarvis 2007).

Following an extensive set of simulations to quantify the sensitivity
of our observations, only those objects that satisfied the following
criteria (equations \ref{magcrit} - \ref{selection}) are considered
candidate Lyman $\alpha$ haloes:

\begin{align}
& 19.5 < nb_{AB} < 23.5 \label{magcrit}\\
& EW_{obs} > 200{\rm \AA}\\
& FWHM_{nb} > 2.6{\rm \asec.}
\label{selection}
\end{align}

\noindent All objects matching these criteria were inspected by eye to
remove contaminants and data artefacts such as high proper motion
objects, amplifier cross-talk and so on from the sample (again, see
Smith \& Jarvis {\it in prep} for more details.). In this paper we
focus on one particular halo which is identified with the source AMS05
(Mart\'inez-Sansigre \etal, 2005).

\subsection{AMS05: Imaging from our survey}\label{intro:ams05}

Mart\'inez Sansigre \etal\ (2005) introduced a sample of type-2
obscured QSOs to constrain the QSO fraction (i.e. the fraction of
objects which are unobscured -- i.e. type-1 -- QSOs). This sample of
QSOs was chosen to satisfy the following selection criteria, designed
to pick out the elusive type-2 QSO population at $z \sim 2$:

\begin{itemize}
\item S$_{\rm 24\mu m} > 300\mu$Jy
\item S$_{\rm 3.6\mu m} \leq 45\mu$Jy
\item 350$\mu$Jy $\leq$ S$_{\rm 1.4 GHz} \leq 2$mJy
\end{itemize}

\noindent These criteria select galaxies that are intrinsically
luminous at 24$\mu$m, a result of bright reprocessed hot-dust emission
indicative of the presence of an AGN. They are also chosen to be
relatively raint at 3.6$\mu$m ensuring that the bright QSO nucleus
does not dominate the shorter wavelengths, i.e. they must be
obscured. The 3.6$\mu$m criterion is also expected to reject sources
with $z \ltsim 1.4$. The 1.4 GHz criteria were designed to remove
the more luminous radio-loud (e.g. FRII-type; Fanaroff \& Riley 1974)
galaxies from the sample, whilst ensuring that the radio emission was
caused by an AGN (i.e. that they are QSOs, rather than starburst
galaxies). These criteria proved highly successful at picking out
type-2 QSOs at $z \sim 2$; the final sample included 21 candidates
(the sample selection is discussed in much greater detail in
e.g. Mart\'inez-Sansigre \etal, 2006a).

\begin{figure}
\centering \includegraphics[width=0.85\columnwidth, clip=y,
  angle=0]{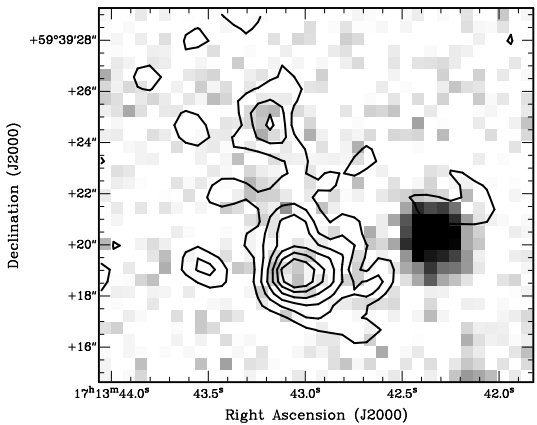}
\caption{AMS05 in g' band, with the continuum-subtracted
  Lyman-$\alpha$ halo overlaid as a contour map. In this frame North
  is up, with East to the left, and the location of the host galaxy's
  peak radio emission coincides with the peak in the Lyman-$\alpha$
  emission. The halo is extended over at least 10\asec; with the
  original discovery spectrum of Mart\'inez-Sansigre \etal\ (2006a)
  being taken with the slit orientated approximately along the
  East-West direction at a position angle of 98$^\circ$ East of North,
  the huge extent of the Lyman-$\alpha$ halo was not detected. The knot
  of ``Lyman-$\alpha$ emission'' to the North-West of the adjacent
  continuum source is probably due to imperfect continuum
  subtraction. The contours correspond to between 23 \& 93\%\ of the
  peak counts per pixel in the continuum subtracted narrow-band image,
  separated at intervals of 10\%.}
\label{FLSm030_nboverlay}
\end{figure}

One of the galaxies that satisfied our Lyman-$\alpha$ halo selection
criteria was a previously known type-2 QSO, dubbed AMS05. Our broad-
and continuum-subtracted narrow-band observations of AMS05 are
presented in figure \ref{FLSm030_nboverlay} and show that the halo
detected in the narrow-band filter is extended over
$\sim$10~arcsec. The original spectroscopy of AMS05 was carried out at
a position angle of 98$^\circ$, meaning that the vast emission line
structure that our narrow-band imaging reveals was not detected. Due
to the large equivalent width estimated from our photometry (EW$_{obs}
> 1000$\AA), and the extent of the emission line structure, we
obtained new spectroscopic observations of AMS05, this time with a
position angle of 12$^\circ$, aligned along the major axis of the
extended emission.

\subsection{Multi-wavelength data used in this study}
\label{publicdata}

In addition to the narrow-band and Sloan-g' band imaging data taken as
part of our survey (Smith \& Jarvis {\it in preparation}), we located
publicly-available imaging observations of the Spitzer First Look
Survey region to constrain the spectral energy distribution of
AMS05. We use u$^\star$-band data from Shim \etal (2006), the R-band
from Fadda \etal (2004), IRAC data from Lacy \etal (2005), MIPS
24$\mu$m data from Fadda \etal (2006), 610~MHz data from Garn \etal
(2007), and 1.4 GHz data from Condon \etal (2003). The data at 4.9 GHz
are from Mart\'inez-Sansigre et al. (2006b). Additional
publically-available data in the J, H \& K$_s$ bands were obtained
using NIRI on the Gemini-North telescope as part of another program
(GN-2005B-Q-75). Images were co-added from 9 different offset
exposures on a (3 $\times$3) square grid pattern (with offsets of 10
-- 20\asec), and then median combined to get the resulting total
integration times of 300, 280, \& 240 seconds in the J, H \& K bands
respectively. In addition, extremely high resolution European Very
Long Baseline Interferometry Network (EVN) data at 1.7~GHz have also
been obtained (H-R. Kl\"ockner, private communication).

This object was also observed with the Infrared Spectrograph (IRS)
aboard the {\it Spitzer Space Telescope} and with the Max-Planck
Millimetre Bolometer Array (MAMBO-2) instrument at the IRAM 30m
telescope at Pico de Veleta, Spain. The IRS observations are described
in some detail in Mart\'inez-Sansigre et al. (2008). MAMBO-2 has an
effective wavelength of 1.2mm, with a beam FWHM of 10.7\asec. During
these observations, the target was centred on the most sensitive pixel
and standard on-off observations were conducted, varying the wobbler
throw between 32, 35 \& 40\asec\ in azimuth at 2~Hz. The atmospheric
opacity was always $\tau_{230~\rm GHz} < 0.3$, and the sky noise was
medium to low ($\leq 200$~mJy~beam$^{-1}$). Gain calibration was
carried out regularly using Neptune, Uranus or Mars, and monitored
using very bright millimetre sources (typically $\geq 5$~Jy). The data
were reduced using MOPSIC (Zylka, 1998) and the absolute flux scale of
MAMBO-2 is uncertain at the 20\%\ level. Photometry of AMS05 is
presented in table \ref{table:FLSm030_phot}.

\begin{table}
\centering
\caption{Photometry of AMS05. Measurements are given in AB magnitudes
  unless otherwise stated. The He{\sc ii}, Sloan g', R, \& i'-band, J,
  H, K, IRAC \& MIPS flux densities presented below are measured in
  8\asec\ diameter apertures. The MAMBO-2 data have a beam size of
  10.7\asec, while the 610 \& 1.4~GHz flux densities are the
  integrated flux densities from the radio catalogues covering the FLS
  (see text for details) and the 1.7~GHz VLBI data were supplied by
  H-R.~Kl\"ockner.  The errors are quoted to 1$\sigma$; where limits
  are given, they correspond to 3$\sigma$. }
\vspace{0.15cm}
\begin{tabular}{|l|l|}
\hline
Photometric Band & Mag$_{AB}$\slash Flux Density \\
\hline
\hline
u$^\star$ & $> 25.75$ \\
He{\sc ii} & 22.26$^{\mbox{ \tiny +0.06}} _{\mbox{ \tiny -0.06}}$ \\
Sloan g'   & 24.98$^{\mbox{ \tiny +0.79}} _{\mbox{ \tiny -0.34}}$ \\
R & 24.90$^{\mbox{ \tiny +0.48}} _{\mbox{ \tiny -0.33}}$ \\
J & $>$ 21.43 \\
H & 20.30$^{\mbox{ \tiny +0.40}} _{\mbox{ \tiny -0.30}}$ \\
K$_s$ & 20.40$^{\mbox{ \tiny +0.39}} _{\mbox{ \tiny -0.29}}$ \\
3.6$\mu$m & 37.8 $\pm$ 2.5 $\mu$Jy \\
4.5$\mu$m & 65.6 $\pm$ 3.3 $\mu$Jy \\
5.8$\mu$m & 142.2 $\pm$ 13.3 $\mu$Jy \\
8.0$\mu$m & 294.8 $\pm$ 12.2 $\mu$Jy \\
24$\mu$m & 1.43 $\pm$ 0.09 mJy \\ 
1.2mm & $<$1.04 mJy  \\
4.9~GHz & 0.44 $\pm$ 0.05 mJy \\
1.7~GHz (VLBI) & 485.6 $\pm$ 24.2 $\mu$Jy \\
1.4~GHz & 1.04 $\pm$ 0.05 mJy \\
610~MHz & 1.31 $\pm$ 0.08 mJy \\
\hline
\end{tabular}
\label{table:FLSm030_phot}
\end{table}

\subsection{Spectroscopy of AMS05}
\label{sec:ISIS_reduction}

\begin{table*}
\centering
\caption{Properties of the different Lyman-$\alpha$ emission line
  components as a function of their separation from the nucleus,
  defined as the radio position which is coincident with the centre of
  the brightest component of the Lyman-$\alpha$ emission. The
  components of the Lyman-$\alpha$ emission furthest away from the
  nucleus are blue-shifted relative to the central component. The
  apertures used here are 2.4\asec\ in length along the 1.5\asec\ wide
  slit.}
\begin{tabular}{ccccc}
\hline
Core separation & Component Flux & Component centre & \multicolumn{2}{|c|}{Component FWHM} \\
(\asec) & (erg cm$^{-2}$ s$^{-1}$) & (\AA) & (\AA) & (km s$^{-1}$) \\
\hline
\hline
0.0 & 2.04 $\times\ 10^{-16}$ & 4681 & 17.3$\pm$3.0 & 1110 $\pm$ 190 \\
2.7 & 7.10 $\times\ 10^{-17}$ & 4675 & 14.2$\pm$3.1 & 910 $\pm$ 200 \\
6.0 & 6.42 $\times\ 10^{-17}$ & 4673 & 12.6$\pm$4.0 & 810 $\pm$ 260 \\
\hline
\end{tabular}
\label{table:spectroscopydata}
\end{table*}

Our spectroscopic observations were taken using the ISIS double-beamed
spectrograph on the William Herschel Telescope on the nights of July
12th \& 15th, 2007. The red arm of ISIS uses the new highly red
sensitive REDPLUS CCD, a 4k$\times$2k array of 15.0$\mu$m pixels,
while the blue arm uses a thinned EEV 4k$\times$2k 13.5$\mu$m pixel
CCD. These observations used the red and blue arms with the R158R and
R300B gratings, for exposure times of 4$ \times$900 and 2$\times$1800
seconds respectively using a 1.5\asec\ slit orientated at a position
angle of 12.0$^\circ$ East of North along the long axis of the
extended emission. The red-arm data were binned up to (3 $\times$ 2),
whilst the blue data were binned (2 $\times$ 2). This resulted in a
spectral resolution of 12.0\AA\ in the red, and 8.5\AA\ for the blue
arm.

The data were de-biassed, illumination- and
flatfield-corrected. Wavelength solutions were derived from
observations of Cu-Ar and Cu-Ne emission line lamps with the same
setup as for the science frames. Flux calibration was derived from
observations of the spectrophotometric standard star
BD+28$^\circ$~4211.

\section{Results}

\subsection{The Halo}
\label{sec:halo}

Although AMS05 is detected in the Sloan-g' band image (figure
\ref{FLSm030_nboverlay}, centre), the observed equivalent width is at
least EW$_{\rm obs} >$ 1000\AA, however this value is uncertain due to
the bright Lyman-$\alpha$ emission dominating the continuum emission
in the broad band image. The flux emitted as Lyman-$\alpha$ falling in
the slit is $3.54 \pm 0.50 \times 10^{-16} {\rm ~erg ~cm}^{-2}{\rm
  ~s}^{-1}$, giving a luminosity of $8.41 \pm 1.19 \times 10^{43} {\rm
  ~erg~s}^{-1}$ (2.19 $\pm\ 0.31 \times 10^{10}$\Lsolar). Our
photometric luminosity estimate of 1.50 $\pm 0.09 \times 10^{44}$ erg
cm$^{-2}$ s$^{-1}$ (3.89 $\pm 0.23 \times 10^{10}$\Lsolar, which is
measured in an 8\asec\ aperture) suggests that slit losses account for
$\sim$ 44\% of the total flux.

\begin{figure}
\begin{centering}
\vspace{0.5cm}
\hspace{0.90cm}\subfigure{\includegraphics[width=0.681\columnwidth]{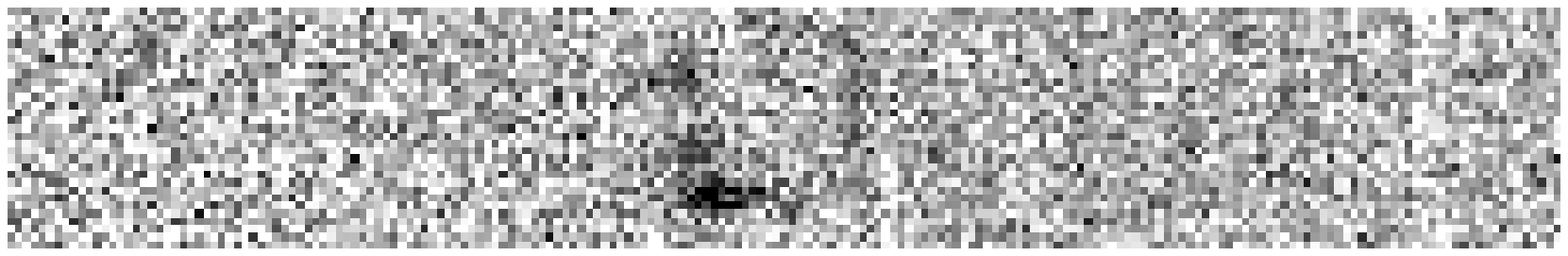}}\\
\vspace{-1.78cm}\subfigure{\includegraphics[width=0.85\columnwidth]{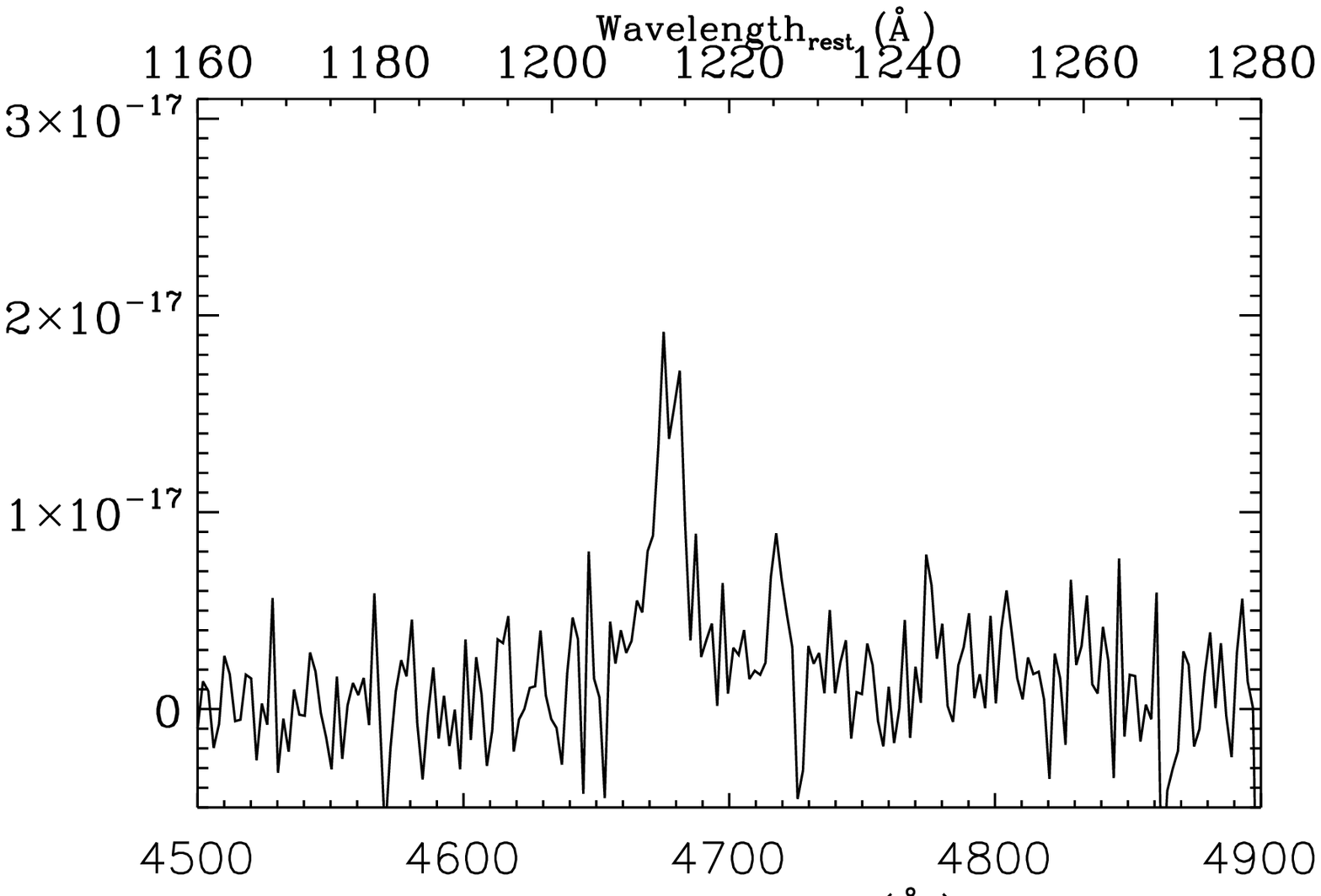}}\\
\caption{One- and two-dimensional spectra of AMS05. The asymmetric
  line profile, high equivalent width, large spatial extent and the
  absence of other emission lines lead us to identify this emission
  line as Lyman-$\alpha$. Three components of the Lyman-$\alpha$ emission
  can be seen in the 2D spectrum (top), in which they are spread over
  $\sim 9$\AA, which corresponds to approximately 580~km~s$^{-1}$. The
  spatial extent of the structure is $>$ 10\asec, which corresponds to
  $\sim80$ kpc at this redshift. The one-dimensional spectrum
  displayed here corresponds to the sum of all the individual
  components (i.e. the extraction aperture was chosen to include all
  of the flux in the two-dimensional spectrum).}
\label{FLSm030_spectra}
\end{centering}
\end{figure}
\begin{figure}
\begin{centering}
\includegraphics[width=0.85\columnwidth]{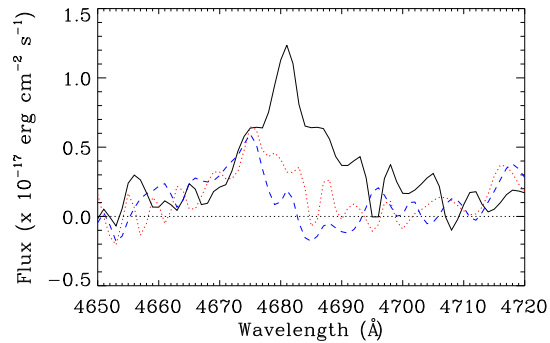}
\caption{Extracted spectra corresponding to the three components of
  the Lyman-$\alpha$ apparent in the two-dimensional spectrum visible
  in figure \ref{FLSm030_spectra} and detailed in table
  \ref{table:spectroscopydata}. Each spectrum has been extracted
  through a 2.4\asec\ aperture centred on positions 0.0, 2.7 and
  6.0\asec\ away from the nucleus, coloured black, red and blue
  respectively (solid, dotted and dashed lines). The nucleus is
  coincident with the radio centroid and the main component of the
  Lyman-$\alpha$ emission. The components of the Lyman-$\alpha$
  emission further away from the nucleus are blueshifted relative to
  the centre; this may be due to either a high-velocity in\slash
  outflow or foreground absorption around the nucleus. }
\label{extracted_spectra}
\end{centering}
\end{figure}

The Lyman-$\alpha$ emission shows a remarkable velocity structure,
apparent in the two-dimensional spectra presented in figure
\ref{FLSm030_spectra}. The Lyman-$\alpha$ emission is split into at
least three different components in the spectrum, with peaks offset
from one another by $\sim 580$~km~s$^{-1}$. Table
\ref{table:spectroscopydata} and figure \ref{extracted_spectra} show
details of spectra extracted in three separate apertures as a function
of their distance from the host galaxy and radio core of the
Lyman-$\alpha$ emitting structure (which coincide with the peak in the
Lyman-$\alpha$ surface brightness).

Upon inspecting figures \ref{FLSm030_spectra} and
\ref{extracted_spectra}, it is immediately apparent that those
components furthest away from the nucleus appear blue-shifted relative
to the central component; they also show narrower linewidths. These
properties bear a striking resemblance to high redshift radio galaxies
(H$z$RG); these relationships are presented for a sample of 15 H$z$RGs
in van Ojik \etal (1997), and also for the three H$z$RGs discussed in
Villar-Mart\'in \etal\ (2007a).

The Lyman-$\alpha$ emission is extended over at least 10\asec\ in both
our narrow-band imaging (figure \ref{FLSm030_nboverlay}), and in our
2D spectrum (figure \ref{FLSm030_spectra}). The spatial extent --
which equates to $\sim$ 80kpc at $z=2.85$ -- may be due to inflow or
outflow of gas surrounding the AGN; this is very difficult to say for
certain. However, for the more powerful radio galaxies, sensitive
polarisation measurements favour the infall scenario (e.g. Humphrey et
al., 2007). To carry out this sort of analysis here would require much
higher sensitivity radio observations with full polarization
properties.

The two most separate components of the Ly$\alpha$ emission line are
offset from one another by approximately 580~km~s$^{-1}$ in
velocity. If we na\"ively interpret this as Keplerian motion, then we
can estimate the mass of the halo as being $\sim$7.6 $\times 10^{11}$
\Msolar. Of course, due to the resonant nature of the Lyman-$\alpha$
photons, it is also quite possible that the regions between the knots
of Lyman-$\alpha$ emission are sites of foreground absorption, or
scattering of Lyman-$\alpha$ photons away from the line of sight (see
e.g. van Ojik et al. 1997; Jarvis et al. 2003; Wilman et al. 2004 \&
2005).

As well as the optical-infrared SED discussed in section
\ref{sed_optir} below, the loosely conical structure of the
Lyman-$\alpha$ emitting region suggests that a central ionizing source
is more likely to be the cause of ionization than any of the other
commonly suggested methods of ionizing a Lyman-$\alpha$ halo of
similar luminosity (see e.g. Smith \& Jarvis, 2007), which are more
likely to appear approximately spherical.

\subsection{The Host galaxy}
\label{sec:host}

\subsubsection{Optical-Infrared SED}
\label{sed_optir}

To constrain the SED of AMS05, we make use of some of the
multi-wavelength data over the FLS field. To analyse the spectral
energy distribution of AMS05, we used a combination of template AGN
and simple stellar population models.  For template AGN models, we
used the Siebenmorgen template AGN SEDs from
\url{http://www.eso.org/~rsiebenm/agn_models/} (see e.g. Siebenmorgen,
\etal, 2004, for a recent review), as well as the mean QSO SED from
Elvis et al. (1994) with a variety of Milky Way model dust
obscurations between $A_V = 0$ and 10 following Pei et al (1992). The
Siebenmorgen template AGN SEDs are the result of self-consistent
radiative transfer simulations with central heating sources. 


The Elvis et al. (1994) mean QSO SED is empirical, based on the
observations of a large sample of radio quiet QSOs. The original
authors noted that the dispersion in their mean SEDs for the
observations of individual objects was around a decade at all
wavelengths. For the simple stellar populations, we made use of the
Bruzual \& Charlot (2003) stellar models. We used simple stellar
populations of ages 10, 30, 50, 100, 200, 400, 600, 1000, 1400, 1800
\& 2200 Myr.

The value of the $\chi^2$ goodness-of-fit parameter was calculated for
each linear combination of stellar population and AGN template, with
the model combination corresponding to the lowest value of $\chi^2$
chosen as our best fit in this regime.

\begin{figure}
\centering
\includegraphics[width=0.95\columnwidth]{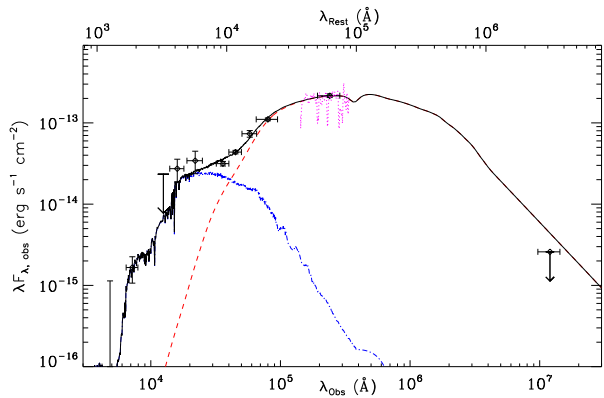}
\caption{Spectral energy distribution of AMS05. The best fit model
  (overlaid in black) consists of a linear combination of a simple
  stellar population from Bruzual \& Charlot (2003) of age 1.4Gyr, a
  Salpeter IMF and solar metallicity (shown in blue), with an AGN
  component from Elvis (1994) with 5.5 magnitudes of visual extinction
  (shown in red). The SED is consistent with the IRS spectroscopic
  observations of Mart\'inez-Sansigre et al. (2008), shown in
  yellow. The evidence for the presence of an AGN within this halo is
  convincing. Whilst the 1.2mm limit appears to disagree with the
  model spectrum, Elvis et al. (1994) noted that there was an order of
  magnitude's dispersion about the mean SED value for individual
  sources (and the SED itself is merely a linear extrapolation between
  their far infrared and radio fluxes in the Elvis et al. sample), and
  so should not be a matter of concern.}
\label{SED}
\end{figure}

To accurately calculate the value of the $\chi^2$ parameter, we
removed the Lyman-$\alpha$ flux from the measurement of the Sloan-g'
band magnitude since Lyman-$\alpha$ haloes are not included in the
Bruzual \& Charlot (2003), Elvis (1994) or Siebenmorgen models. The
presence of such profuse Lyman-$\alpha$ emission would, left unchecked,
produce artificially large measurements for the photometry of these
objects since Lyman-$\alpha$ falls near the centre of the Sloan-g' band
filter's transmission function at the redshift studied here. This
subtraction is reflected in the large error bar in the Sloan-g' band
data point in figure \ref{SED}, since the Lyman-$\alpha$ emission
dominates the Sloan-g' band flux.

The object itself is associated with detections in all bands with the
exception of $u^\star$, J, and 1.2mm. AMS05 was originally selected by
Mart\'inez-Sansigre \etal\ (2005) to have a mid-IR\slash radio
spectral energy distribution consistent with a type 2 high redshift
QSO; our studies are consistent with the presence of an obscured AGN.

The optical-infrared SED of AMS05 is shown in figure \ref{SED}; as
mentioned in Mart\'inez-Sansigre \etal\ (2005) the combination of a
high S\slash N detection at 24$\mu$m ($\sim6.2~\mu$m in the emitted
frame), faint flux at 3.6$\mu$m (restframe 9350 \AA -- z' band), and a
moderate radio luminosity of $L_{1400 MHz} \sim 10^{32.4}$ erg
s$^{-1}$~Hz$^{-1}$ sr$^{-1}$ ensures that the host galaxy to this
Lyman-$\alpha$ halo contains a powerful AGN rather than a
starburst. The AGN is thought to be obscured by the torus invoked by
unified schemes to explain the varying properties of AGN with
orientation with respect to the line of sight (see e.g. Urry \&
Padovani, 1995). This is in agreement with our SED fitting; we find
that the best fit to the SED of AMS05 is a combination of a 1.4~Gyr
old simple stellar population and an obscured Elvis (1994) AGN
component with $A_V$ = 5.5. The best fit composite SED is also shown
in figure \ref{SED}.

Whilst the 1.2mm non-detection would appear to be at odds with our
best-fit SED, the uncertainties associated with the 1.2mm observations
(for which the flux calibration has an uncertainty of $\sim$20\%), in
tandem with the large inherent scatter in QSO SEDs mean that this
apparent discrepancy can be easily explained. It is quite possible
that AMS05 has less cool dust than expected from the median SED from
Elvis et al. (1994), which was merely a linear extrapolation between
the far-infrared and radio wavelengths in any case.

We can use the normalisations of the SED components to estimate the
stellar mass and AGN luminosity. For the stellar SED we derive a mass
of 4.4 $\pm\ 0.3 \times 10^{10}$\Msolar, similar in mass to powerful
radio galaxies at the same redshift (e.g. Jarvis et al. 2001, Willott
\etal, 2003), and a bolometric AGN luminosity of 3.4 $\pm\ 0.2 \times
10^{13}$\Lsolar, with the latter number being derived assuming that
$L_{\rm bol} = 10 \times \nu L_{\nu}$ at 8 $\mu$m (Elvis et al., 1994,
Richards et al. 2006), and assuming the Milky-Way dust extinction law
of Pei (1992). Note that the uncertainty on the bolometric correction
alone could be as high as 50\%.

\subsubsection{Results from millimetre-wavelength observations}

AMS05 was not detected to a 3$\sigma$ limit of 1.04~mJy at 1.2mm. To
constrain the star formation rate due to cold dust mass within the
halo, we consider the 1.2mm observed flux to be due to any of three
different templates; a {\it grey body}, and templates based on M82 and
Arp220. For the grey body template, we assume typical emissivity index
and dust temperature from Omont et al. (2003 - $\beta = 1.5$, $T =
45~K$), and for all models we adopt the Kennicutt (1998) calibration
of SFR to far-IR luminosity (between 8 and 1000$\mu$m). Using these
assumptions we find that the star formation rate derived assuming grey
body emission in the far infrared is $< 650 \Msolar$~yr$^{-1}$ (or $<
250 \Msolar$~yr$^{-1}$ if we assume a lower dust temperature, $T =
35~K$). The M82 and Arp220 templates suggest higher limits on the star
formation rates of $\ltsim$700 \& 950 \Msolar~yr$^{-1}$, respectively;
it is clear that there may be a vigorous amount of obscured star
formation residing within AMS05 that we do not observe in our optical
SED (figure \ref{SED}).

\subsubsection{Radio SED}


\begin{figure}
\centering \includegraphics[width=0.95\columnwidth]{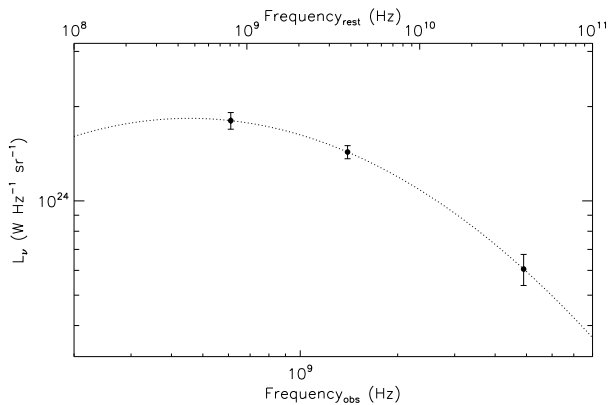}
\caption{Radio properties of AMS05; the spectral energy distribution
  appears to turn over around $\sim$1.8~GHz in the rest-frame. Whilst
  in radio-loud FRI\slash FRII objects this may be interpretted as a
  young radio source, this is not the case for AMS05, due to its radio
  quiet nature.} 
\label{gps}
\end{figure}

The radio SED of AMS05 is displayed in figure \ref{gps}, overlaid with
a polynomial of the form $\log_{\mbox{\tiny 10}}S_\nu =
\Sigma^{\mbox{\tiny 2}} _{\mbox{\tiny $i$=0}}a_i \log_{10} \nu^i$
(following Jarvis \& Rawlings, 2001). With a second order polynomial
and three data points the best fit is exact; the radio SED of AMS05
peaks at 455~$\pm$ 275~MHz in the observed frame (where the errors
have been derived analytically, and are consistent with a set of
100,000 Monte-Carlo simulations of our observational results). This
corresponds to $\sim$1.7 $\pm$ 1.0~GHz in the rest frame.

Gigahertz peaked spectrum radio sources are usually young, radio-loud
objects, whose relativistic jets have not escaped the confines of the
host galaxy (e.g. O'Dea et al., 1998). However, despite the high radio
luminosity (it lies close to the Fanaroff Riley break of $L_{178} \sim
2 \times 10^{24}$ W Hz$^{-1}$~sr$^{-1}$), we will now argue that AMS05
contains a luminous {\em radio quiet} AGN, and therefore we can draw
no parallels between typical GPS sources and the source in question
here.

Quasars are considered to be radio-loud or radio-quiet based on the
ratio of their radio to optical luminosity, $R$, with $R=10$ generally
considered to be the dividing line (Kellerman \etal, 1989). The same
calculation can be made for a type-2 AGN like AMS05 if we consider its
unobscured optical luminosty, derived from our SED fit. We derive
$R=8.5\pm0.7$ which places AMS05 in the radio-quiet regime, albeit as
a bright example of this class. This is further supported by the EVN
observations, which reveal that $\sim$50\% of the flux at
$\sim$1.4~GHz is emitted on scales $<$220~pc, indicating the absence
of the bulk of the radio emission on kpc scales. It
is highly unlikely that the difference between the EVN and VLA fluxes
is due to AGN variability; whilst a factor of two in variation is
typical of blazars, they demonstrate strong beamed continuum and\slash
or broad emission lines, neither of which is observed in AMS05.

\begin{figure}
\centering \includegraphics[width=0.85\columnwidth]{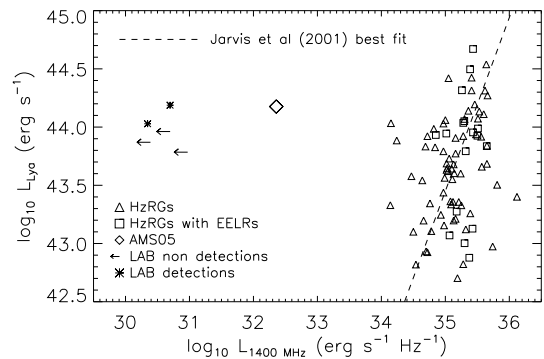}
\caption{Lyman-$\alpha$ versus rest-frame 1400 MHz luminosity for a
  selection of high redshift galaxies known to possess Lyman-$\alpha$
  haloes (Eales \etal, 1993, McCarthy \etal, 1993; de Breuck \etal,
  2000; Steidel \etal, 2000; Francis \etal, 2001; Jarvis \etal, 2001,
  Maxfield \etal, 2002, Reuland \etal, 2003, Villar-Mart\'in \etal,
  2003, Dey \etal, 2005; Smith \& Jarvis, 2007, Villar-Mart\'in \etal,
  2007). Radio luminosities at 1400 MHz have been derived assuming
  power law SEDs of the form $S_\nu \propto \nu^{-\alpha}$, where
  $\alpha = 0.8$, except for AMS05, which uses the gigahertz peaked
  radio SED. HzRGs from de Breuck \etal, 2000 have been included in
  order to better demonstrate that AMS05 (diamond) does not belong to
  this class of object.}
\label{radio_lyman}
\end{figure}

Whilst the Lyman-$\alpha$ emission bears great resemblance to the
EELRs around H$z$RGs, the above arguments indicate that AMS05 is an
intrinsically different class of source. This can most clearly be seen
in figure \ref{radio_lyman}; using the polynomial fit to extrapolate
the radio luminosity of the LAB host galaxy to 151 MHz in the rest
frame we find that even though the luminosity in Lyman-$\alpha$
emission is similar to the median HzRGs drawn from the 3CRR, 7CRS, 6CE
and 6C$\star$ samples in Jarvis \etal\ (2001), the luminosity at radio
frequencies is several orders of magnitude lower. 

This source appears in many ways to be very similar to the
low-redshift radio-quiet quasar E1821+643 (Blundell \& Rawlings 2001),
which exhibits quasar-like optical emission but also has FRI-type
radio emission over 100~kpc scales. Blundell \& Rawlings explain this
as a precessing jet, possibly due to a binary black hole system at the
centre of the galaxy. This precession could result in an FRI
morphology, with around 50 per cent of the radio luminosity coming
from the core, but the detectability of the extended emission becomes
very difficult due to the reduced surface brightness on large
scales. Future low-frequency telescopes such as LOFAR would be ideally
placed to map out such extended structures around such radio-quiet
AGN.


\begin{figure}
\centering
\includegraphics[width=0.95\columnwidth, clip=y]{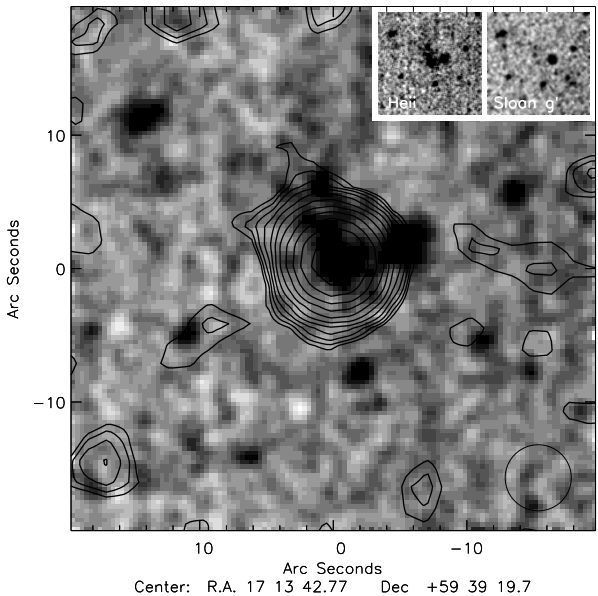}
\caption{AMS05 in Sloan-g' band data with the 1400MHz VLA data
  overlaid; the beam size is shown to the bottom right of the
  frame. The insets to the top right indicate the \Heii, and Sloan-g'
  band thumbnails to enable easier comparison with the 1400MHz data.}
\label{AMS05_overlay}
\end{figure}

\subsubsection{Photon Number counts}

We demonstrate that the AGN generates enough photons to ionize the
observed Ly$\alpha$ halo, using a photon number counts argument,
similar to that described by Neugebauer et al., (1980). By using the
median unobscured QSO SED from Elvis (1994), scaled to match our
measured flux at 24$\mu$m (where the effects of obscuration by dust
are negligible and the QSO dominates the monochromatic luminosity), we
can estimate the number of photons emitted by the naked source before
the effects of obscuration along our line of sight are taken into
account. We find that the nucleus emits around 1.2 $\times 10^{57}$
ionizing photons per second, which is $>2$ orders of magnitude more
than that which is emitted as Lyman-$\alpha$ photons. There is no
doubt that the active nucleus is capable of ionizing the
Lyman-$\alpha$ halo, although a starburst component could easily
contribute as is postulated for HzRGs (e.g. Binette et al. 2006;
Villar-Mart\'in et al. 2007b), however this contribution is
constrained by the MAMBO-2 observations.

\section{Conclusions}
\label{haloconclusions}

We have discovered the first highly-extended Lyman-$\alpha$-emitting
halo around a radio-intermediate type-2 obscured QSO. The halo, which
is extended over $\sim$80~kpc, is similar to those haloes observed
around high-redshift radio galaxies, and the more quiescent
Lyman-$\alpha$ blobs, with a luminosity estimated from our photometry
of 1.50 $\pm\ 0.09\times 10^{44}$\fluxunits (3.89 $\pm\ 0.23 \times
10^{10}$\Lsolar).  By analogy with the high-redshift radio galaxies,
where polarisation data help determine the source orientation on the
sky, we interpret the velocity offset and decrease in velocity
dispersion toward the edge of the halo as evidence for infall rather
than outflow, although again this would need deeper radio observations
to determine directly.

The existence of Lyman-$\alpha$ ``fuzz'' around QSOs at $z \sim 3$ was
predicted by Haiman \& Rees (2001), and Alam \& Miralda--Escud\'e
(2002) - with suggested angular sizes $\gtsim 2$\asec and
0.1-1.0\asec, respectively. Whilst one example of extension on small
angular scales was presented in Francis \& McDonnell (2006), this new
discovery is the first detection of a QSO hosting an extensive
Lyman-$\alpha$ halo of which the authors are aware.

The halo is associated with a Gigahertz-peaked radio SED, although
since AMS05 is radio quiet, this does not imply that the radio source
is young. The most likely cause of ionization of the large extended
Lyman-$\alpha$ emission for this halo is photoionization from the
central radio-quiet AGN. However, we cannot rule out that at least
part of the excess in Ly$\alpha$ emission may arise from other
sources, such as an episode of star formation (e.g. Binette et al.,
2006) or cooling radiation (e.g. Fardal et al. 2001; Dijkstra et
al. 2006a,b).

AMS05 appears similar to the low-redshift radio-quiet quasar E1821+643
which has large-scale FRI radio struture but an optically powerful AGN
at its centre. Blundell \& Rawlings (2001) suggest that the lack of
bright emission on large scales in this source is due to a precessing
jet caused by a binary black hole system, which would result in lower
luminosity emission on the kpc scale. Future observations with the
next generation of radio telescopes could detect such emission around
optically powerful but sub-luminous radio sources.

\section*{Acknowledgments}
The authors wish to thank Hans-Rainer Kl\"ockner for sending
unpublished VLBI data. DJBS wishes to thank the UK STFC for a
PDRA. MJJ acknowledges the support of a Research Council UK
fellowship. The Isaac Newton and William Herschel Telescopes are
operated on the island of La Palma by the Isaac Newton Group in the
Spanish Observatorio del Roque de los Muchachos of the Instituto de
Astrofisica de Canarias. Based in part on observations obtained at the
Gemini Observatory, which is operated by the Association of
Universities for Research in Astronomy, Inc., under a cooperative
agreement with the NSF on behalf of the Gemini partnership: the
National Science Foundation (United States), the Science and
Technology Facilities Council (United Kingdom), the National Research
Council (Canada), CONICYT (Chile), the Australian Research Council
(Australia), CNPq (Brazil) and SECYT (Argentina). IRAM is supported by
INSU/CNRS (France), MPG (Germany), and IGN (Spain).

\bsp 

\label{lastpage}

\end{document}